\documentclass[a4paper,prx,aps,twocolumn,preprintnumbers,amsmath,amssymb,superscriptaddress]{revtex4-2}

\usepackage[utf8]{inputenc}
\usepackage{graphicx}
\usepackage{times}
\usepackage{amsmath}
\usepackage{amsfonts}
\usepackage{amssymb}
\usepackage{siunitx}

\usepackage{color}
\definecolor{red}{rgb}{1,0,0}
\definecolor{blue}{rgb}{0,0,1}
\definecolor{black}{rgb}{0,0,0}

\newcommand{\ket}[1]{\mbox{$| #1 \rangle$}}

\newcommand{\DR}{\ensuremath{\mathrm{DR}}}

\usepackage{color}
\definecolor{darkblue}{rgb}{0,0,0.6}

\definecolor{red}{rgb}{0.6,0,0}

\definecolor{green}{rgb}{0,0.6,0}

\definecolor{grey}{rgb}{0.7,0.7,0.7}

\begin{document}

\title{Magneto-optical chirality in a coherently coupled exciton-plasmon system}

\author{Samarth Vadia}
\altaffiliation{These authors contributed equally to this work}
\def\LMU{Fakult\"at f\"ur Physik, Munich Quantum Center, and Center for NanoScience (CeNS), Ludwig-Maximilians-Universit\"at M\"unchen, Geschwister-Scholl-Platz 1, 80539 M\"unchen, Germany}
\affiliation{\LMU}
\affiliation{Munich Center for Quantum Science and Technology (MCQST), Schellingtr. 4, 80799 M\"unchen, Germany}  
\affiliation{attocube Systems AG, Eglfinger Weg 2, 85540 Haar, Germany}

\author{Johannes Scherzer}
\altaffiliation{These authors contributed equally to this work}
\affiliation{\LMU}

\author{Kenji Watanabe}
\affiliation{Research Center for Functional Materials, National Institute for Materials Science, 1-1 Namiki, Tsukuba 305-0044, Japan}

\author{Takashi Taniguchi}
\affiliation{International Center for Materials Nanoarchitectonics, 
National Institute for Materials Science, 1-1 Namiki, Tsukuba 305-0044, Japan}

\author{Alexander H\"ogele}
\affiliation{\LMU}
\affiliation{Munich Center for Quantum Science and Technology (MCQST), Schellingtr. 4, 80799 M\"unchen, Germany}  

\date{\today}

\begin{abstract}
Chirality is a fundamental asymmetry phenomenon, with chiral optical elements exhibiting asymmetric response in reflection or absorption of circularly polarized light. Recent realizations of such elements include nanoplasmonic systems with broken mirror symmetry and polarization-contrasting optical absorption known as circular dichroism. An alternative route to circular dichroism is provided by spin-valley polarized excitons in atomically thin semiconductors. In the presence of magnetic fields, they exhibit an imbalanced coupling to circularly polarized photons and thus circular dichroism. Here, we demonstrate that polarization-contrasting optical transitions associated with excitons in monolayer WSe$_2$ can be transferred to proximal plasmonic nanodisks by coherent coupling. The coupled exciton-plasmon system exhibits magneto-induced circular dichroism in a spectrally narrow window of Fano interference, which we model in a master equation framework. Our work motivates exciton-plasmon interfaces as building blocks of chiral metasurfaces for applications in information processing, non-linear optics and sensing.
\end{abstract}

\maketitle

Direct band gap and reduced dielectric screening in semiconducting monolayer transition metal dichalcogenides (TMDs) \cite{Mak2010,Splendiani2010} give rise to tightly bound excitons \cite{Wang2018coll} with sizable light-matter interactions that facilitate efficient coupling to dielectric or plasmonic systems \cite{Liu2015,Dufferwiel2015,Schneider2018}. Capitalizing on the large oscillator strength of TMD excitons and the flexibility of combining them with plasmonic structures, recent examples of coupled exciton-plasmon systems include realizations in the strong and weak coupling regimes \cite{Liu2016,Kleemann2017, Kern2015,Lee2015}. While the former is characterized by the formation of exciton-plasmon polaritons, the latter is distinguished by Fano-type interference spectra, as discussed in early work on various exciton-plasmon coupled systems \cite{Wiederrecht2004, Zhang2006, Govorov2010}. In this framework, the dipolar selection rules of spin-valley polarized excitons in TMD monolayers \cite{Xiao2012,Cao2012,Mak2012,Zeng2012,Sallen2012} provide a route to chiral optical phenomena, as the valley degeneracy can be lifted by magnetic field to induce spectrally imbalanced coupling to left- and right-handed circularly polarized photons \cite{Li2014Valley,Aivazian2015,Srivastava2015,Wang2015b,Koperski2018}. This opto-valleytronic feature of TMD monolayer excitons has been utilized to demonstrate chiral effects such as directional coupling of light in silver nanowires on WS$_2$ \cite{Gong2018}, spatial separation of valley-polarized excitons by silver nanogroove arrays \cite{Sun2019}, or second-harmonic generation of circularly polarized photons in gold-WS$_2$ metasurfaces \cite{Hu2019}.

Here, we study magneto-optical characteristics of an exciton-plasmon metasurface based on a WSe$_2$ monolayer and gold (Au) nanodisks. We elucidate the effect of Fano interference as a function of exciton-plasmon spectral resonance detuning in the weak coupling regime \cite{Fano1961,Miroshnichenko2010,Lee2015,Abid2017,Sun2018,petric2021tuning}, and study both experimentally and theoretically the polarization properties of the coherently coupled system in the presence of external magnetic fields. Remarkably, the coupled system exhibits magnetic circular dichroism that is distinct from the characteristics of the fundamental valley-polarized exciton transition in monolayer WSe$_2$. The resulting chiral behavior of the Fano-coupled metasurface manifests in the form of a spectral window with polarization dependent reflectivity in an otherwise broadband opaque medium. The observations are substantiated by a master equation analysis with excellent quantitative agreement with experimental findings. Our work provides insight into the underlying coherent interference phenomena and can serve as a guideline to the design of exciton-plasmon metasurfaces with optical chirality in the visible spectral range.

\begin{figure}[t]
\centering
\includegraphics[scale=1]{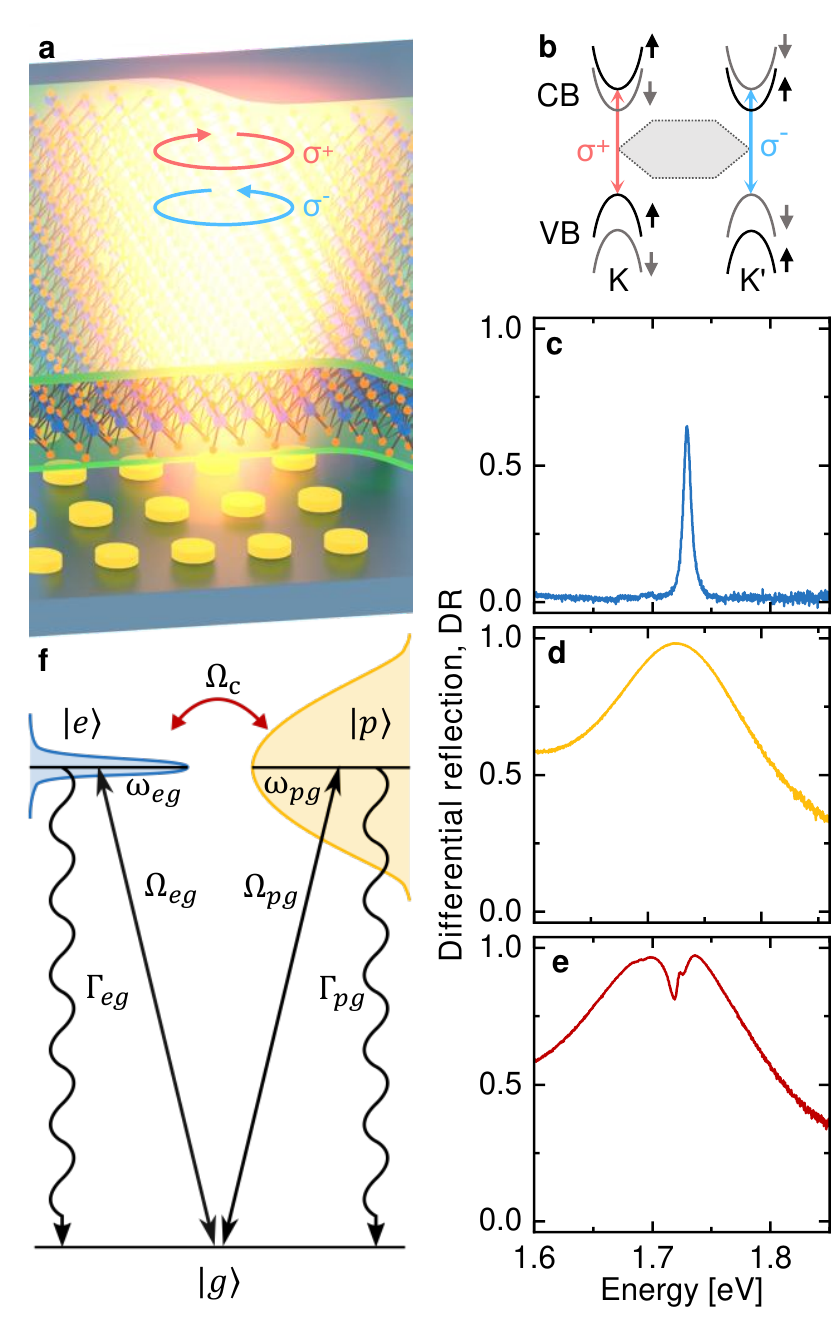}
\caption{\textbf{Fano interference in a coupled exciton-plasmon system.} \textbf{a}, Schematic illustration of monolayer WSe$_2$ encapsulated in hBN, placed on gold nanodisks and probed with circularly polarized light. \textbf{b}, Band structure schematics of monolayer WSe$_{2}$ at two opposite corners of the hexagonal Brillouin zone, with $\sigma^{+}$ ($\sigma^{-}$) circularly polarized optical transitions between spin-up (spin-down) polarized states at the K (K') valleys of the conduction band (CB) and valence band (VB). \textbf{c}, \textbf{d}, and \textbf{e}, Differential reflection spectra of monolayer WSe$_2$, gold nanodisk, and the coupled system, respectively. \textbf{f}, Energy levels of the coupled system with \ket{g}, \ket{e} and \ket{p} denoting the ground, exciton and plasmon state, respectively, and the corresponding exciton and plasmon optical transition frequencies $\omega_{eg}$ and $\omega_{pg}$, Rabi frequencies $\Omega_{eg}$ and $\Omega_{pg}$, and radiative decay rates $\Gamma_{eg}$ and $\Gamma_{pg}$, as well as the exciton-plasmon coupling strength $\Omega_{c}$.}
\label{fig1}
\end{figure}

The exciton-plasmon interface was fabricated by encapsulating a monolayer WSe$_2$ in hexagonal boron nitride (hBN) and placing the resulting heterostructure on top of a plasmonic Au nanodisk array on a SiO$_2$/Si substrate, as illustrated in Fig.~\ref{fig1}a (see Methods for sample details). The sample features regions of encapsulated WSe$_2$ monolayer, Au nanodisk arrays, as well as regions where both elements are combined in vertical proximity. To characterize the optical responses of the bare exciton and plasmon systems, and the regime of their coupling, we performed differential reflection spectroscopy at cryogenic temperatures (see Methods for experimental details). The corresponding spectra are shown in Fig.~\ref{fig1}c, d and e, where differential reflection $\DR=(\mathrm{R_{sub}}-\mathrm{R})/(\mathrm{R_{sub}})$ was measured relative to the reflection $\mathrm{R_{sub}}$ of the SiO$_2$/Si substrate. The coupled system can be modelled in a three-level system framework with relevant states and rates shown in Fig.~\ref{fig1}e. The DR spectra in Fig.~\ref{fig1}c, d and e are representative for the excitation from the ground state $\ket{g}$ to the exciton state $\ket{e}$ with resonance and Rabi frequency $\omega_{eg}$ and $\Omega_{eg}$ and decay rate $\Gamma_{eg}$ (Fig.~\ref{fig1}c), the optical excitation to the plasmon state $\ket{p}$ with resonance and Rabi frequencies $\omega_{pg}$ and $\Omega_{pg}$ and decay rate $\Gamma_{pg}$ (Fig.~\ref{fig1}d), and the simultaneous excitation of the interacting exciton-plasmon system with coherent coupling constant $\Omega_c$ (Fig.~\ref{fig1}e).

The band structure of WSe$_2$ monolayer is shown schematically in Fig.~\ref{fig1}b, together with polarization-contrasting optical transitions of the fundamental exciton X$^0$ in the $K$ and $K'$ valleys of the hexagonal Brillouin zone. The transitions are degenerate at zero magnetic field, resulting in a single Lorentzian peak in the DR spectrum of Fig.~\ref{fig1}b at $\hbar \omega_{eg}~\simeq 1.723$~eV. The full-width at half-maximum (FWHM) linewidth of the exciton transition $\hbar\Gamma_{eg} \simeq 8$~meV is substantially smaller than the plasmon linewidth of $\hbar\Gamma_{pg} \simeq 180$~meV obtained from the region of a bare Au nanodisk array in Figure~\ref{fig1}d with resonance energy at $\hbar \omega_{pg} = 1.72$~eV. Due to variations in the dielectric environment of the nanodisk array, the plasmon resonance energy $\hbar\omega_{pg}$ varies in the range of $100$~meV (see Supplementary Section 1 for the description of gold nanodisk arrays). In our studies, this variation beneficially provides position-dependent spectral energy detuning $\delta=\hbar\omega_{pg}-\hbar\omega_{eg}$ of the plasmon resonance energy with respect to the exciton resonance which has negligible variations across the sample. 

\begin{figure}[t]
\centering
\includegraphics[scale=1]{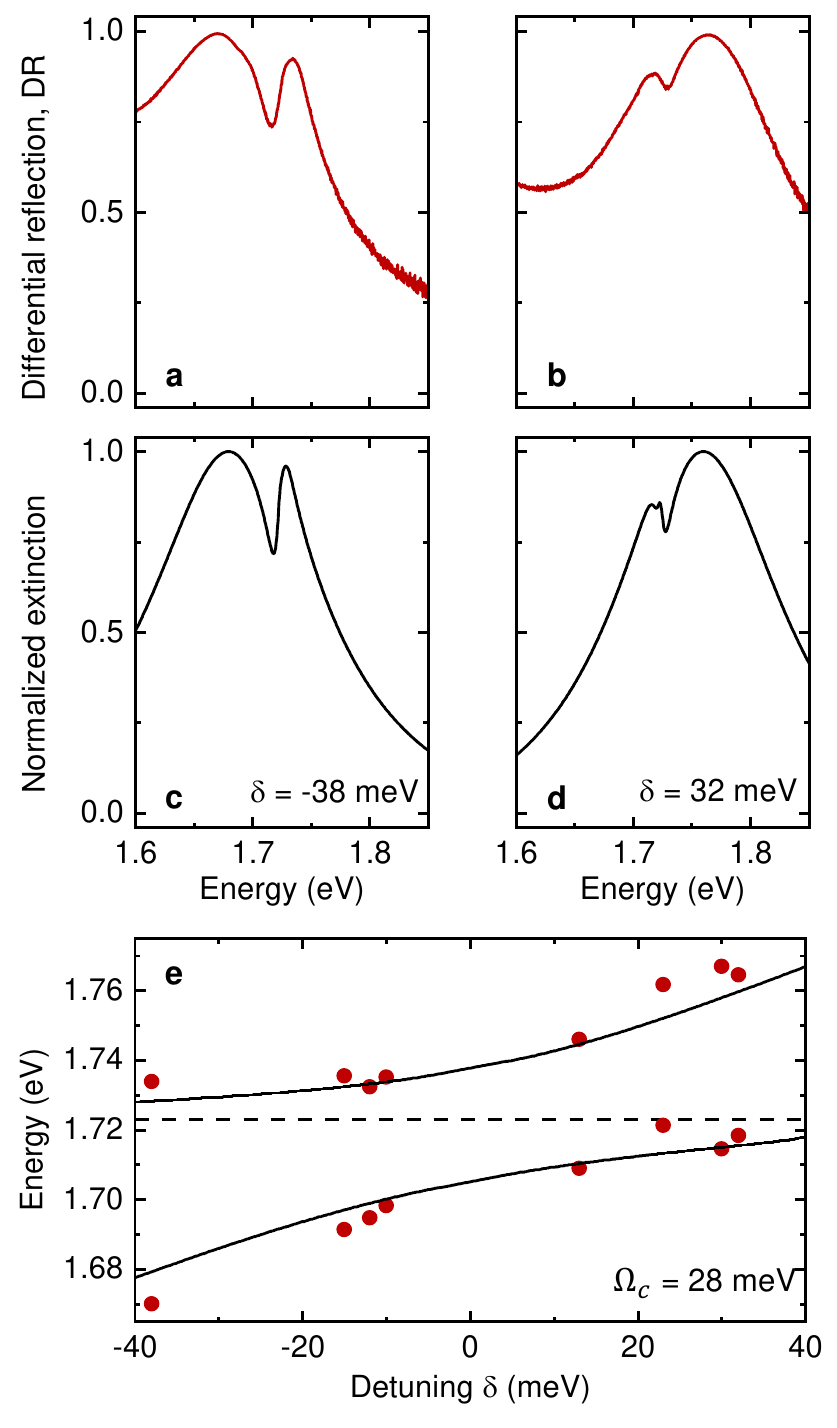}
\caption{\textbf{Exciton-plasmon Fano interference in experiment and theory}. \textbf{a} and \textbf{b}, Normalized differential reflection spectra of the coupled system for $-38$ and $32$~meV energy detuning from the exciton-plasmon spectral resonance condition. The spectra were recorded on two positions of the nanodisk array with different plasmon energies $\hbar\omega_{pg}$ (determined from Lorentzian fits) for a weakly varying exciton energy $\hbar\omega_{eg}$. \textbf{c} and \textbf{d}, Respective Fano model spectra with exciton-plasmon coupling strength $\hbar\Omega_c=28$~meV. \textbf{e}, Evolution of the exciton-plasmon coupling as a function of resonance energy detuning $\delta$, with red data points corresponding to plasmonic array regions with different $\hbar\omega_{pg}$ and respective model results (black lines) obtained with the bare exciton energy $ \hbar \omega_{eg} = 1.723$~eV (dotted line) and $\hbar\Omega_{c}=28$~meV.}
\label{fig2} 
\end{figure}

The spectrum of the coupled system in Fig.~\ref{fig1}e is characterized by a Fano interference lineshape  \cite{Fano1961,Miroshnichenko2010,Limonov2017,Li2018} with a narrow reflection dip in the broad plasmonic extinction peak. A closer inspection of the spectrum reveals an additional peak superimposed on the reflection dip, which we ascribe to the contribution from uncoupled excitons that are located within the optical spot yet sufficiently far away from plasmonic nanodisks. To interpret the resulting lineshape, we inspected system realizations with different resonance detunings at different spatial positions of the interfaced array. Two representative $\mathrm{DR}$ spectra for negative and positive resonance detuning $\delta$ are shown in Fig.~\ref{fig2}a and b, respectively. In both spectra, the position of the dip remains essentially constant due to small variation in the exciton energy across the sample, which also holds for the uncoupled exciton peak inside the Fano dip. Due to the asymmetric character of the Fano interference, however, the overall spectral shape is strongly modified at different resonance conditions. 

We model this intricate optical response in the framework of two coherently coupled oscillators using a classical light field interacting with exciton and plasmon dipolar excitations. The dipole moments of the respective optical transitions are given as the imaginary components of the quantum coherence obtained from the master equation analysis (see Supplementary Section 2 for theoretical modeling of the extinction spectrum). All main parameters of the system including the decay rates $\Gamma_{eg}$, $\Gamma_{pg}$ and the Rabi frequencies $\Omega_{eg}$, $\Omega_{pg}$ were determined from experiments on bare system components. To quantify the coupling strength $\Omega_{c}$, we plot the spectral position of the two maxima enclosing the dip in the Fano spectra as a function of detuning $\delta$ in Fig.~\ref{fig2}e, with their energy separation reproduced by the theoretical model for a coupling strength $\hbar\Omega_{c} = 28$~meV, as shown by solid lines in Fig.~\ref{fig2}e. With this coupling, our model yields the normalized extinction spectra shown in Fig.~\ref{fig2}c and d for resonance detunings $\delta=-38$ and $32$~meV as extracted from the spectra in Fig.~\ref{fig2}a and b. All features of the optical response are reproduced with good agreement by the theoretical model.

\begin{figure}[t]
\centering
\includegraphics[scale=1]{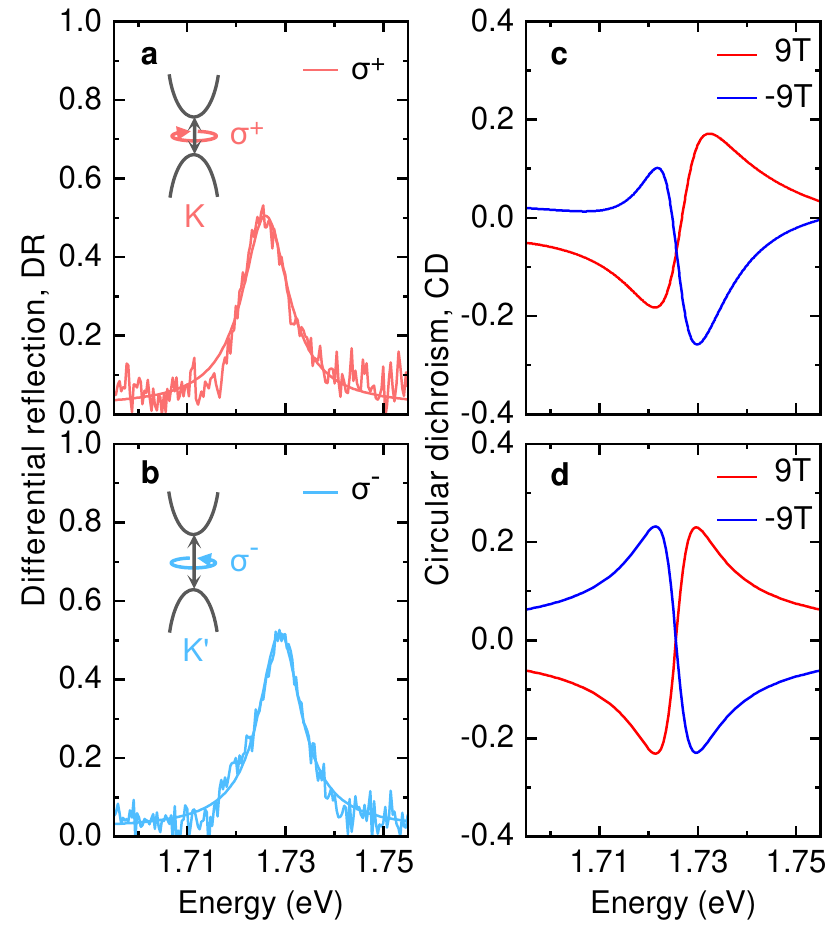}
\caption{\textbf{Magnetic circular dichroism of the exciton transition.} \textbf{a} and \textbf{b}, Valley-selective differential reflection spectra at $9$~T for $\sigma^{+}$ and $\sigma^{-}$ polarized excitation, respectively (solid lines show Lorentzian fits). \textbf{c} and \textbf{d}, Circular dichroism of the exciton transition at $9$ (red) and $-9$~T (blue) from experiment and theory, respectively.}
\label{fig3}
\end{figure}

With this understanding of the Fano interference phenomena, we elucidate in the following the magneto-optical response of the coupled exciton-plasmon system. First, we quantify the degree of circular dichroism (CD) associated with the exciton valley Zeeman effect in monolayer WSe$_2$ \cite{Aivazian2015,Srivastava2015,Wang2015b,Koperski2018}. In the presence of an out-of-plane magnetic field of $9$~T, circularly polarized $\mathrm{DR}$ spectra of Fig.~\ref{fig3}a and b reveal two exciton resonances associated with $K$ and $K'$ transitions that couple to $\sigma^{+}$ and $\sigma^{-}$ polarized light, respectively. The valley Zeeman splitting of $2.1$~meV corresponds to the exciton Land\'e factor with an absolute value of $4$ as expected from previous experiments \cite{Srivastava2015,Wang2015b,Koperski2018} and theory \cite{Wozniak2020,Foerste2020exciton,Deilmann2020,Xuan2020}.

The polarization-contrasting response of the two valleys is quantified by $\mathrm{CD}$, which calculates as $\mathrm{CD} = (\mathrm{DR}^- - \mathrm{DR}^+) / (\mathrm{DR}^- + \mathrm{DR}^+)$, where $\mathrm{DR}^+$ and $\mathrm{DR}^-$ are the the $\sigma^+$ and $\sigma^-$ polarized $\mathrm{DR}$ spectra respectively. Figure~\ref{fig3}c shows the $\mathrm{CD}$ at $9$~T as red solid line, where $\mathrm{DR}^+$ and $\mathrm{DR}^-$ are the Lorentzian fits to the $\sigma^+$ and $\sigma^-$ polarized $\mathrm{DR}$ spectra shown as solid lines in Fig.~\ref{fig3}a and b, respectively. It shows a reversal in polarity around the resonance energy of the exciton at $0$~T with maximum $\mathrm{CD}$ of $\sim 20\%$. The $\mathrm{CD}$ obtained from the corresponding experiments at $-9$~T, shown in Fig.~\ref{fig3}c in blue, is reversed in sign for the entire spectral range (deviations from the mirror-symmetry around the exciton energy at zero field stem from sample inhomogeneities sampled by spatial displacements in magnetic field over a range of $18$~T). All main features of the spectra are captured by our model $\mathrm{CD}$ spectra in Fig.~\ref{fig3}d obtained from extinction.

\begin{figure}[t]
\centering
\includegraphics[scale=1]{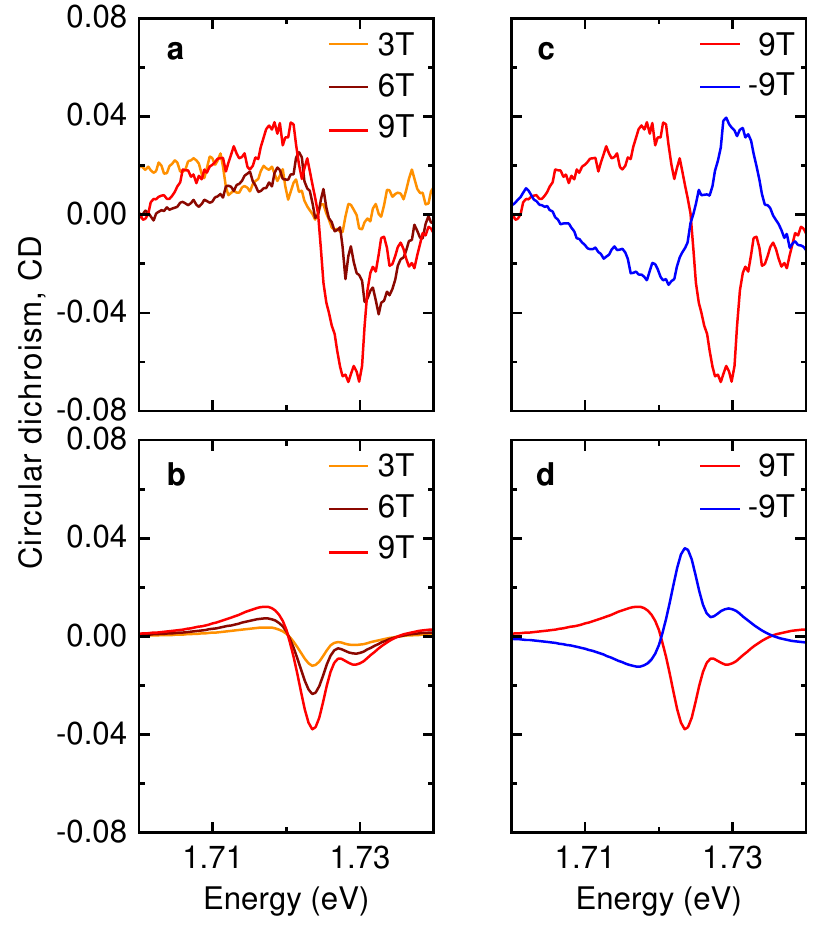}
\caption{\textbf{Magnetic circular dichroism of the coupled exciton-plasmon system.} \textbf{a} and \textbf{b}, Experimental and theoretical circular dichroism spectra of the coupled exciton-plasmon system at $3$, $6$ and $9$~T. \textbf{c} and \textbf{d}, Experimental and theoretical circular dichroism spectra in magnetic fields of $9$ (red) and $-9$~T (blue).}
\label{fig4}
\end{figure}

Next, we study the polarization-dependent optical response of the interface region in the presence of a magnetic field. The valley Zeeman effect of the bare exciton is imprinted on the coupled exciton-plasmon system in an intricate way and manifests as polarization-dependent reflectance. Our theory captures the experimentally observed features of the $\mathrm{CD}$ spectra and their evolution with the magnetic field, as evidenced by comparing experimental and theoretical data shown in Fig.~\ref{fig4}a and b and Fig.~\ref{fig4}c and d, respectively. The evolution of $\mathrm{CD}$ with increasing magnetic fields of $3$, $6$ and $9$~T, recorded in the region of the Fano interference with $-15$~meV resonance detuning, is shown in Fig.~\ref{fig4}a. The magneto-induced dichroism becomes increasingly pronounced with increasing magnetic field as for the bare WSe$_2$ monolayer. Notably, the spectra in Fig.~\ref{fig4}c show a sign reversal of the $\mathrm{CD}$ response for the coupled system in comparison to the bare exciton case in Fig.~\ref{fig3}c. While the monolayer features a peak in the exciton DR spectrum, the coupled exciton-plasmon spectrum is characterized by a narrow dip around the exciton resonance as a result of Fano interference, leading to reversed $\mathrm{CD}$ response and pronounced asymmetry for finite exciton-plasmon detunings. In the calculated spectra of Fig.~\ref{fig4}d, the asymmetry manifests in the form of an additional dip in the $\mathrm{CD}$ spectra, in qualitative agreement with the experimental spectra in Fig.~\ref{fig4}c. The contribution of uncoupled excitons within the optical focal spot with a reversed sign of $\mathrm{CD}$ explains this observation. A maximum $\mathrm{CD}$ of up to $\sim 7~\%$ is achieved in the coupled exciton-plasmon system, indicating a significant transfer of the opto-valleytronic exciton features onto the coupled system. Consistently, the $\mathrm{CD}$ spectra exhibit a sign reversal at magnetic fields of $\pm 9$~T.

Our observation of magneto-optical effects in a coherently coupled exciton-plasmon system and their detailed quantitative understanding provide a pathway to design ultrathin metasurfaces for chiral spectral filtering, which could be also exploited to control nonreciprocal phenomena with magnetic field for  uni-directional flow of circularly polarized photons as required for information transfer in quantum networks \cite{Kimble2008,Reiserer2015}. An obvious way to create a permanent chiral exciton-plasmon metasurface is to utilize layered ferromagnets that induce sizable exciton valley Zeeman splittings of several meV, equivalent to external magnetic fields well above $10$~T \cite{Zhong2017,Zhao2017,Seyler2018,Norden2019,Ciorciaro2020}. Furthermore, spectrally narrow response from TMD excitons in the limit of atomically thin mirror \cite{Bettles2016,Zeytinogu2017,Shahmoon2017,Back2018} would yield spectrally sharp Fano reflectance or windows of transparency in a broad extinction response. As such, spectral regions with perfect destructive quantum interference and negative refractive index equivalent to electromagnetically induced transparency could be realized for chiral slow light and information storage \cite{Boller1991,Hau1999,Lukin2001}.

\vspace{11pt}
\noindent \textbf{Methods}\\
\noindent The sample was fabricated by depositing monolayer WSe$_{2}$ (HQ Graphene) embedded in high-quality hBN (NIMS) onto a gold nanodisk array. The array was fabricated by standard electron-beam lithography and gold evaporation on a Si/SiO$_{2}$ substrate. All measurements were carried out at cryogenic temperatures. The data in Fig.~\ref{fig1} and Fig.~\ref{fig2} were recorded in a helium bath cryostat at $4.2$~K, whereas the magnetic field measurements of Fig.~\ref{fig3} were performed in a closed-cycle magneto-cryostat (attocube systems, attoDRY1000) at $3.5$~K. White-light reflection spectroscopy was performed using a halogen lamp (Ocean Optics, HL-2000) or a supercontinuum laser (NKT, SuperK Extreme EXR-4) focused to a spot of $\sim$ \SI{1}{\micro\meter} diameter in a home-built confocal microscope equipped with cryogenic nano-positioners (attocube systems, ANP100 and ANP101 series) and micro-objectives (attocube systems, LT-APO/VISIR/0.82 or LT-APO/LWD/NIR/0.63). The reflected light was spectrally dispersed by a monochromator (Roper Scientific, Acton SP2500 or Acton 300i) and detected by a CCD (Roper Scientific, Spec~10:100BR/LN or Andor, iDus 416).

\vspace{11pt}
\noindent \textbf{Acknowledgements}\\
\noindent The authors thank A.~O.~Govorov for fruitful discussions, and P.~Altpeter and C.~Obermayer for continuous support in the clean room. This research was funded by the European Research Council (ERC) under the Grant Agreement No.~772195 as well as the Deutsche Forschungsgemeinschaft (DFG, German Research Foundation) within the Priority Programme SPP~2244~2DMP (Project No. 443405595) and the Germany's Excellence Strategy Munich Center for Quantum Science and Technology (MCQST) EXC-2111-390814868. S.\,V. acknowledges funding from the European Union's Horizon 2020 research and innovation programme under the Marie Sk{\l}odowska-Curie Grant Agreement Spin-NANO No.~676108. K.\,W. and T.\,T. acknowledge support from JSPS KAKENHI (Grant Numbers 19H05790, 20H00354 and 21H05233). 

\vspace{11pt}
\noindent \textbf{Corresponding authors}\\ 
S.\,V. (samarth.vadia@physik.lmu.de), J,\,S. (johannes.scherzer@physik.lmu.de), and A.\,H. (alexander.hoegele@lmu.de).


\end{document}


\title{Magneto-optical chirality in a coherently coupled exciton-plasmon system \\ \vspace{24pt} \Large{Supporting Information} \vspace{24pt} }

\author{Samarth Vadia}
\altaffiliation{These authors contributed equally to this work}
\def\LMU{Fakult\"at f\"ur Physik, Munich Quantum Center, and Center for NanoScience (CeNS), Ludwig-Maximilians-Universit\"at M\"unchen, Geschwister-Scholl-Platz 1, 80539 M\"unchen, Germany}
\affiliation{\LMU}

\affiliation{Munich Center for Quantum Science and Technology (MCQST), Schellingtr.~4, 80799 M\"unchen, Germany}

\affiliation{attocube Systems AG, Eglfinger Weg 2, 85540 Haar, Germany}

\author{Johannes Scherzer}
\altaffiliation{These authors contributed equally to this work}
\affiliation{\LMU}

\author{Kenji Watanabe}
\affiliation{Research Center for Functional Materials, National Institute for Materials Science, 1-1 Namiki, Tsukuba 305-0044, Japan}

\author{Takashi Taniguchi}
\affiliation{International Center for Materials Nanoarchitectonics, 
National Institute for Materials Science, 1-1 Namiki, Tsukuba 305-0044, Japan}

\author{Alexander H\"ogele}
\affiliation{\LMU}
\affiliation{Munich Center for Quantum Science and Technology (MCQST), Schellingtr.~4, 80799 M\"unchen, Germany}  

\date{\today}

\maketitle
\newpage
\noindent 

\section*{Supporting Note 1: Gold nanodisk arrays}
The metasurface discussed in this work consists of a monolayer WSe$_{2}$ encapsulated in hexagonal boron nitride (hBN) and placed on a gold (Au) nanodisk array. The individual metallic nanoparticles support plasmon resonaces localized to the size of the particle. They are characterized by electric field enhancement in their direct proximity under optical excitation, which we use to couple exciton and plasmon resonances. The spectral position of the plasmon resonance depends on the shape and size of the particles, as well as on their dielectric environment. 

\begin{figure*}[t]
\centering
\includegraphics[scale=1.0]{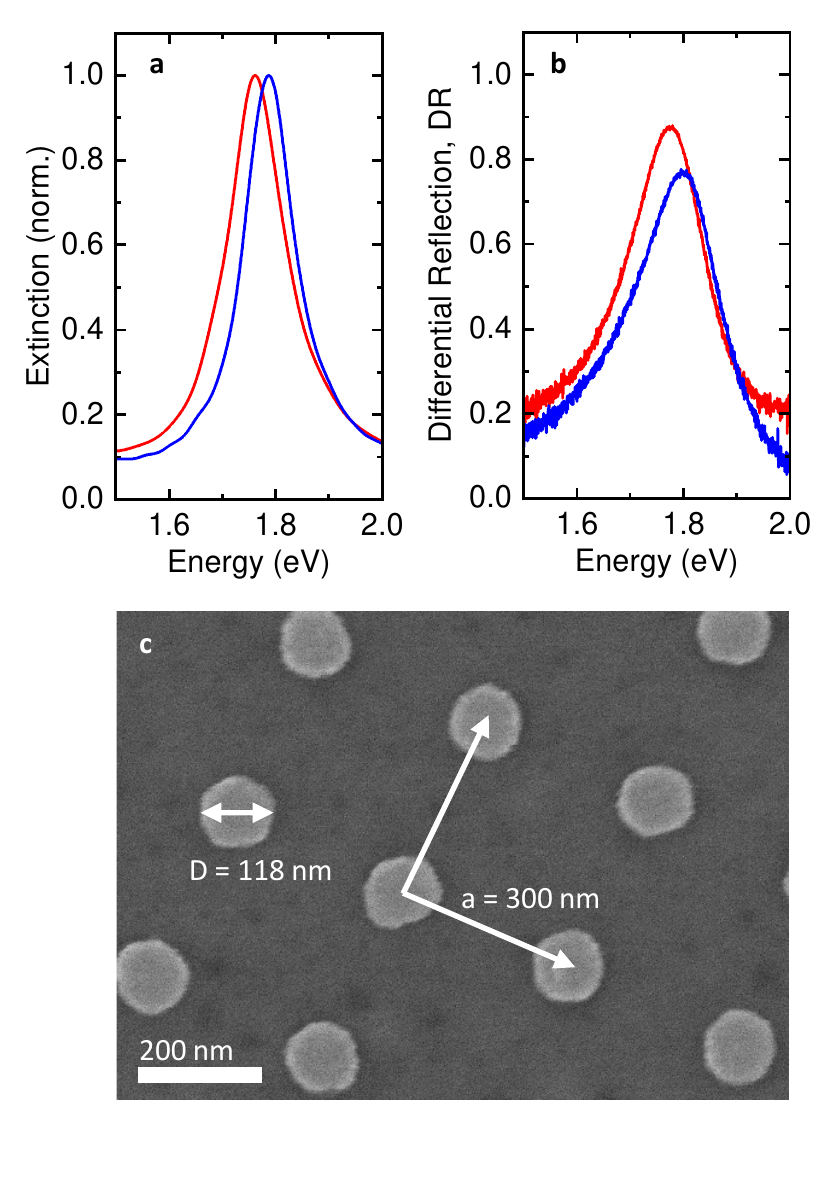}
\caption{\textbf{a}, Extinction spectra of two different gold nanodisk arrays on a silver/SiO$_{2} $ substrate obtained from FDTD simulation. The reflected (R) and transmitted (T) intensities were evaluated at the both ends of the computational cell, and normalized to the signal of the plane wave source. The extinction is defined as $\mathrm{1-R-T}$. Both arrays have a lattice constant $a$ of $300$~nm and height $h$ of $15$~nm. Red and blue spectra correspond to diameters $D$ of $114$ and $106$~nm,  respectively. \textbf{b}, Differential reflection spectra of two gold nanodisk arrays fabricated via e-beam lithography on a silver/SiO$_{2} $ substrate. The geometrical parameters correspond to the ones used for the simulation in \textbf{a}. \textbf{c}, Scanning electron micrograph of an array section of the sample investigated in the main text with mean nanodisk diameter $D$ of $118$~nm.}
\label{figSEM}
\end{figure*} 

The design of plasmonic nanodisk arrays respects the following basic considerations. First, the plasmon resonance of the Au nanodisks has to match the energy of the exciton transition in WSe$_{2}$ to enable near-resonance coherent coupling. Second, the TMD monolayer has to be placed in close proximity to the Au nanodisks for the near-field enhancement of the exciton-plasmon interaction. Third, the height of the nanodisks should be as small as possible to avoid exciton localization in the form of quantum dots due to the variation in lateral strain in the TMD monolayer  \cite{Palacios2017,Branny2017}. And fourth, finally, the number of nanodisks within the focal spot under optical excitation should be large to increase the effective coupling strength.

\begin{figure*}[t]
\centering
\includegraphics[scale=1.0]{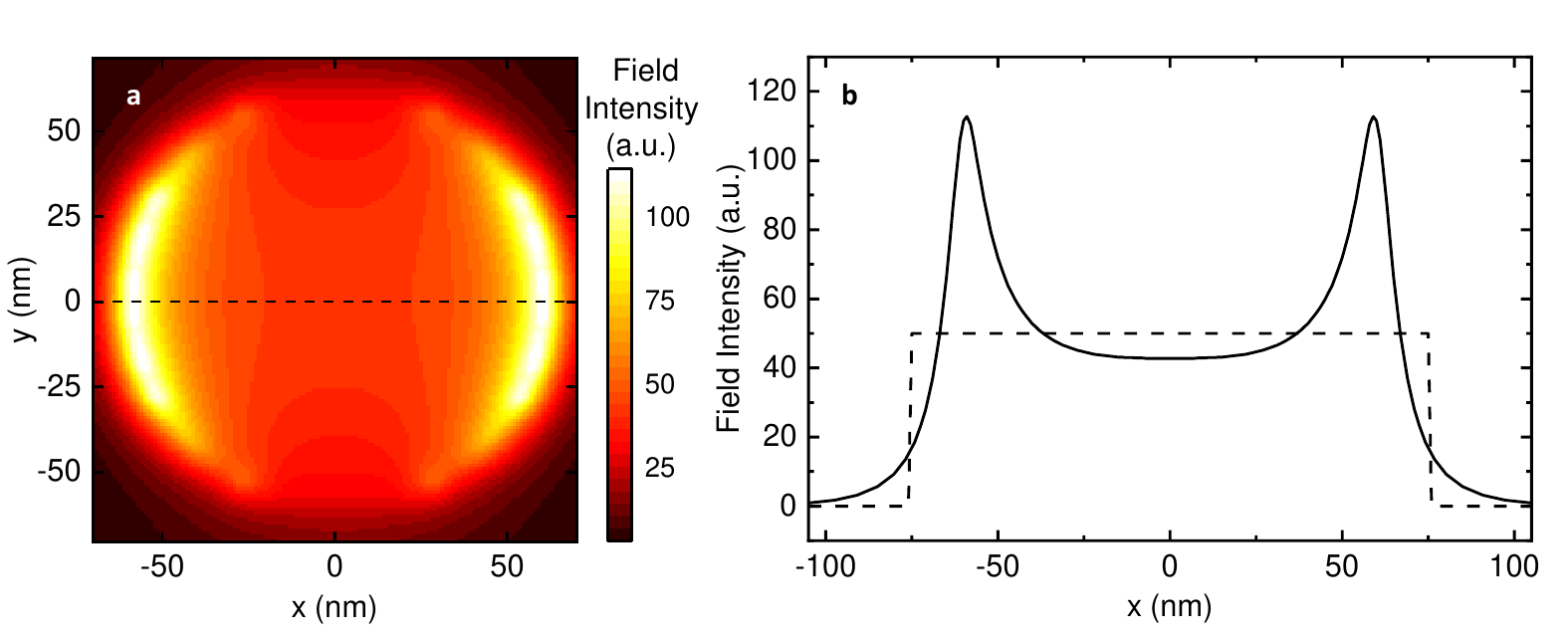}
\caption{\textbf{a}, Simulated field intensity at the cross-section area of the FDTD computational cell $6$~nm above an Au nanodisk (at the position of the monolayer) at the plasmon resonance. The intensity is normalized to the intensity of the excitation plane wave source, thus effectively determining the intensity enhancement factor for the near-field of the plasmonic nanodisk. \textbf{b}, One-dimensional line cut (dashed line in \textbf{a}) of the intensity cross-section area shown in \textbf{a}. The dashed line at field intensity of $50$ represents the approximation of the intensity within the overlap area $A_{c}$ of the monolayer and the plasmonic near-field enhancement shown in \textbf{a}.}
\label{figFDTD}
\end{figure*}

Following these considerations, we performed finite-difference time-domain (FDTD) simulations with a commercial software (Lumerical) to find the optimal geometrical parameters for our system. A SiO$ _{2} $ coated silver mirror was used as substrate for the nanostructure both in simulations and in actual samples. A thin layer with refractive index $n=1.76$ simulates the hBN interface at the top surface of the nanodisks. The main geometrical parameters for tuning the plasmon resonance of the disks are their diameter and height. With the consideration to minimize the nanodisk height in mind, we fixed the height at $\sim 15$~nm where evaporated gold is expected to form smooth surfaces. The plasmon resonance can be shifted to smaller energies by increasing the disk diameter, while a blue-shift accompanies diameter reduction. Figure~\ref{figSEM}a shows this behavior when simulating the extinction of an optical plane wave propagating through gold nanodisks with different diameters. To enhance the near-field effect, we used a $6$~nm thin bottom hBN layer in our van der Waals heterostack defining the distance between the TMD monolayer and the Au nanodisk array. To estimate the magnitude of near-field enhancement at the monolayer position, we evaluated the intensity of the electric field in FDTD simulation in the monolayer plane as plotted in Fig.~\ref{figFDTD}a. The mean intensity value across the entire area of a nanodisk is shown by the dashed line in the intensity line profile in Fig.~\ref{figFDTD}b and serves as an input parameter for the model of the coupled exciton-plasmon system. The plasmonic response in differential reflection varies within a range of $100$~meV across the metasurface. The shift of the plasmon resonance is significant only in regions where the array was covered by hBN, which indicates that the variation of the dielectric environment, related to inhomogeneities in the distance between the hBN layer and the plasmonic structure, dominates the variation of the plasmon energy.   

Using the simulation results as input parameters, we fabricated gold nanodisk arrays using electron beam lithography. Figure~\ref{figSEM}c shows a scanning electron micrograph of the Au nanodisk array used in the experiments of the main text. The optimized interparticle distance for a maximum density of nanodisks within the periodic array was $a = 300$~nm. We found that smaller array spacings compromised the lift-off process for nanodisk diameters larger than $100$~nm. The resolution of our electron beam lithography process was dependent on the e-beam voltage with a limit of $\sim 10$~nm. This limited resolution determined the range of variations in the disk diameters in an array of otherwise identical fabrication parameters. The variation of individual nanodisk diameters in turn resulted in inhomogeneous broadening of the plasmon resonances measured in differential reflection (see Fig.\ref{figSEM}b) as compared to the results of corresponding FDTD simulations shown in Fig.~\ref{figSEM}a. The experimental full-width at half-maximum (FWHM) linewidth of the plasmon resonances was $\mathrm{FWHM}\simeq 180$~meV, and was used as the plasmon decoherence rate in the model detailed in the following section.

\clearpage

\section*{Supporting Note 2: Coherent interference in the coupled-oscillator model}
To model the coupled system of excitons in WSe$_2$ monolayer and localized surface plasmons in Au nanodisk arrays interacting with a light field of frequency $\omega$, we consider a three-level system comprising \ket{g}, \ket{e} and \ket{p} as the ground, exciton and plasmon state, respectively. The energy separation between the states \ket{g} and \ket{e} and the states \ket{g} and \ket{p} is $\hbar \omega_{eg}$ and $\hbar \omega_{pg}$, respectively. The Rabi frequencies for the respective transitions are $\Omega_{eg}$ and $\Omega_{pg}$. In addition, the coherent coupling between the states \ket{e} and \ket{p} is denoted by $\Omega_{c}$. The Hamiltonian of the system is written as:
\begin{equation}
H=\hbar \quad\left(\begin{array}{ccc} 
0 & \frac{1}{2}\Omega_{eg} e^{i \omega t} & \frac{1}{2}\Omega_{pg} e^{i \omega t} \\
\frac{1}{2} \Omega_{eg} e^{-i \omega t} & \omega_{eg} & \frac{1}{2}\Omega_{c} \\
\frac{1}{2}\Omega_{pg} e^{- i \omega t} & \frac{1}{2}\Omega_{c} & \omega_{pg}
\end{array}
\right).
\label{Hamiltonian}
\end{equation}

To describe the coherent dynamics of the Hamiltonian including the decoherence and dissipation channels, we use the master equation for the density matrix, given by \cite{Loudon2000}:
\begin{equation}
\dot{\rho}=-\frac{i}{\hbar} \left.[ H, \rho\right]+ L\rho,
\label{mastereqn}
\end{equation}
\noindent where the density matrix $\rho$ with matrix elements  $\rho_{ij}$ describes the statistical state of the system. The diagonal matrix elements $\rho_{ij}$ (with $i=j$) are the populations of the states, with $i=g$, $e$ or $p$, and the off-diagonal matrix elements $\rho_{ij}$ (with $i \neq j$) are the coherences between the states $i$ and $j$ with complex conjugate $\bar{\rho}_{ij}=\rho_{ji}$. The second term in the equation, $L \rho$, accounts for the dissipation in the Lindblad form.

First, we apply the rotating-wave approximation (RWA) to remove the time-dependent terms in the Hamiltonian of Eq.~\ref{Hamiltonian}. Using $\rho_{ii}=\widetilde{\rho}_{ii}$, $\rho_{eg}=\widetilde{\rho}_{eg} e^{- i \omega t}$, $\rho_{pg} = \widetilde{\rho}_{pg} e^{- i \omega t}$, and $\rho_{ep}=\widetilde{\rho}_{ep}$, the Hamiltonian in RWA modifies to: 
\begin{equation}
\widetilde{H} =\hbar \quad\left(\begin{array}{ccc} 
0 & \frac{1}{2}\Omega_{eg} & \frac{1}{2}\Omega_{pg} \\
\frac{1}{2}\Omega_{eg} & \Delta_{eg} & \frac{1}{2}\Omega_{c} \\
\frac{1}{2}\Omega_{pg} & \frac{1}{2}\Omega_{c} & \Delta_{pg}
\end{array}
\right),
\label{RWAHamiltonian}
\end{equation}
\noindent where $\Delta_{eg} = \omega_{eg} - \omega$ and $\Delta_{pg} = \omega_{pg} - \omega$ are the respective detunings between the transition frequencies $\omega_{eg}$ and $\omega_{pg}$ and the laser frequency $\omega$. Omitting the tilde superscript in the following for simplicity, we describe the dynamic evolution of the density matrix elements by the master equation in the Lindblad form as:
\begin{equation}
\begin{aligned}
&\dot{\rho}_{gg}=-i\left[\frac{\Omega_{e g}}{2}\left(\rho_{ge}-\rho_{eg}\right)+\frac{\Omega_{pg}}{2}\left(\rho_{pg}-\rho_{gp}\right)\right] + \Gamma_{ee} \rho_{ee} + \Gamma_{pp} \rho_{pp} \\
&\dot{\rho}_{ee}=-i\left[\frac{\Omega_{e g}}{2}\left(\rho_{eg}-\rho_{ge}\right)+\frac{\Omega_{c}}{2}\left(\rho_{pg}-\rho_{gp}\right)\right] -\Gamma_{ee} \rho_{ee} \\ 
&\dot{\rho}_{pp}=-i\left[\frac{\Omega_{pg}}{2}\left(\rho_{gp}-\rho_{pg}\right)+\frac{\Omega_{c}}{2}\left(\rho_{ep}-\rho_{pe}\right)\right] - \Gamma_{pp} \rho_{pp} \\
&\dot{\rho}_{ge}=-i\left[\frac{\Omega_{e g}}{2} (\rho_{ee}- \rho_{gg})-\Delta_{eg} \rho_{ge}+\frac{\Omega_{pg}}{2} \rho_{pe}-\frac{\Omega_{c}}{2} \rho_{gp}\right]- \frac{\Gamma_{eg} \rho_{ge}}{2}\\
&\dot{\rho}_{gp}=-i\left[\frac{\Omega_{pg}}{2}\left(\rho_{pp}-\rho_{gg}\right)+\frac{\Omega_{eg}}{2} \rho_{ep}-\frac{\Omega_{c}}{2} \rho_{ge}-\Delta_{pg} \rho_{gp}\right]- \frac{\Gamma_{pg} \rho_{gp}}{2} \\
&\dot{\rho}_{ep}=-i\left[\frac{\Omega_{e g}}{2} \rho_{gp}-\frac{\Omega_{pg}}{2} \rho_{eg}+\Delta_{eg} \rho_{ep}+\frac{\Omega_{c}}{2}\left(\rho_{pp}-\rho_{ee}\right)-\Delta_{pg} \rho_{ep}\right]
\end{aligned}.
\end{equation}

Dissipation is introduced in the above master equation via population decay rates, $\Gamma_{ee}$ and $\Gamma_{pp}$, and the decoherence rates, $\Gamma_{eg}$ and $\Gamma_{pg}$ \cite{Rand2016}. The population decay rate includes contributions from radiative and non-radiative decay channels, $\Gamma_{ii} = \Gamma_{ii}^{r} + \Gamma_{ii}^{nr}$, respectively. The decoherence rate is related to the population decay rate as $\Gamma_{ig} = \Gamma_{ii} + 2 \gamma_{ig, deph}$ where $\gamma_{ig, deph}$ is the pure dephasing rate. 

The Rabi frequency $\Omega_{ig}$ between the excited state \ket{i} (\ket{i} = \ket{e} or \ket{p}) and the ground state \ket{g} is given by:
\begin{equation}
\Omega_{ig} = \frac{\mu_{ig} \mathcal{E}}{\hbar} = \sqrt{\frac{6 \pi c^2 \Gamma_{ii} P}{\hbar n \omega_{ig}^3 A}}
\end{equation}
\noindent where $\mu_{ig}$ is the dipole moment of the transition, $\mathcal{E}$ is the electric field, $c$ is the speed of light, $\Gamma_{ii}$ is the population decay rate of the state \ket{i}, $P$ is the illumination laser power, $n$ is the refractive index, $\omega_{ig}$ is the transition frequency from the excited state \ket{i} to the ground state \ket{g}, and $A$ is the illumination area. For the transition between the exciton state \ket{e} and the ground state \ket{g}, there are two distinct Rabi frequencies, namely the coupled Rabi frequency $\Omega_{ig,c}$ frequency for excitons in the near-field of plasmonic nanodisks affected by the plasmonic enhancement of the electric field, and the uncoupled Rabi frequency $\Omega_{ig,u}$ of excitons away from Au nanodisk but still excited within the optical focal spot. The Rabi frequency of coupled exciton fraction is related to the Rabi frequency of the uncoupled fraction as $\Omega_{eg,c} = \Omega_{eg,u} \sqrt{F} A_{c}/A_{u}$, where $A_{c}$ is the overlap area between the monolayer and the nanodisk with plasmonic near-field enhancement, $A_{u}$ is the area of uncoupled excitons within the optical focal spot, and $F$ is the enhancement factor of the electric field obtained from the FDTD calculation. 

Solving the set of coupled equations for steady state where $\dot{\rho}_{ij} = 0$, and using population conservation in the three-level system as $\rho_{gg} + \rho_{ee} + \rho_{pp} = 1$, we obtain analytical expressions for the populations and coherences in the coupled system as a function of experimental parameters. The optical response is obtained from the imaginary part of the coherence matrix elements \cite{Gerardot2008}. In our system, the focal spot consists of an exciton-plasmon hybrid region showing coherent interference and also a region with uncoupled excitons in monolayer away from the nanodisk. Therefore, we obtain the extinction $\mathrm{E}$ in the optical response as the sum of the coherence matrix elements of two coupled states and the uncoupled exciton state as follows:
\begin{equation}
\mathrm{E} \; \propto \; \alpha_{e} \frac{\Omega_{eg,c}\Gamma_{ee}^{r}}{\Omega_{eg,c}^{2}+\Omega_{pg}^{2}} \mathrm{Im}(\rho_{ge}) + \alpha_{p} \frac{\Omega_{pg}\Gamma_{pp}^{r}}{\Omega_{eg,c}^{2}+\Omega_{pg}^{2}}\mathrm{Im}(\rho_{gp}) + \alpha_{e} \frac{\Gamma_{ee}^{r}}{\Omega_{eg,uc}} \mathrm{Im}(\rho_{ge,u}),
\end{equation}

\noindent where $\alpha_{e}$ and $\alpha_{p}$ correspond to the maximum on-resonance scattering contrast of the exciton and plasmon transition with radiative decay rates $\Gamma_{ee}^{r}$ and 
$\Gamma_{pp}^{r}$. The respective contrasts $\alpha_{i}$ are obtained as \cite{Karrai2003}:
\begin{equation}
\alpha_i = \frac{3}{2 \pi} \frac{c^{2}}{n^2 \omega_{ig}^2 A}.
\end{equation}
\noindent We note here that the extinction is not quantitative as all reflecting surfaces of the sample are not taken into account to calculate the Rayleigh scattering field based on the procedure detailed in \cite{Karrai2003}. However, the response spectrum is proportional to the imaginary part of the coupled coherences $\rho_{eg}$ and $\rho_{pg}$, and the uncoupled exciton coherence $\rho_{eg,u}$ as a function of frequency, and we obtain correct qualitative description of differential reflection. 

\begin{figure*}[t!]
\centering
\includegraphics[scale=0.98]{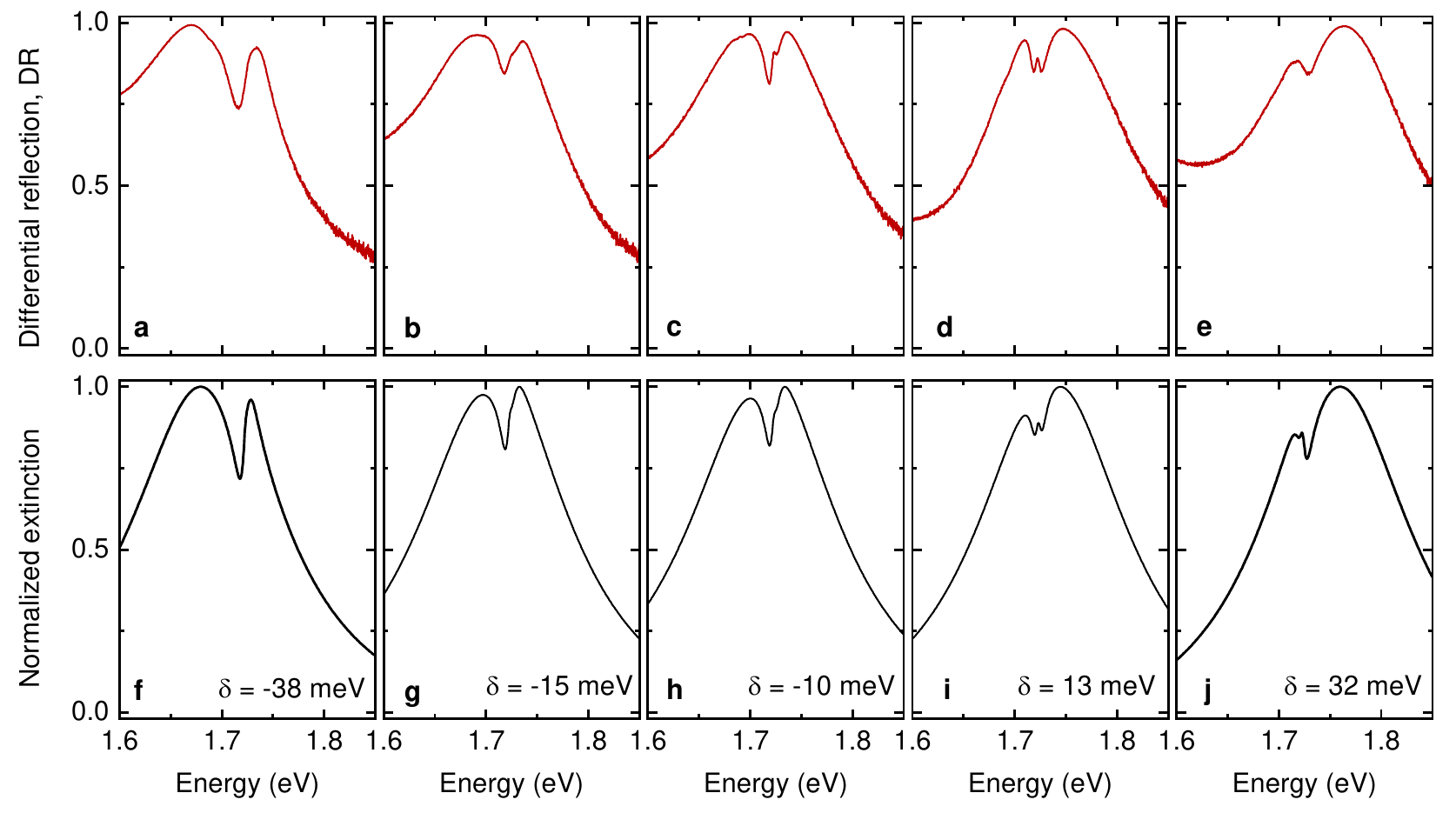}
\caption{\textbf{a -- e}, Differential reflectance spectra of the coupled system for various detunings $\delta=\hbar\omega_{pg} - \hbar \omega_{eg}$. The spectra were recorded on different spatial positions with different plasmon frequencies $\omega_{pg}$ and a weakly varying exciton frequency $\omega_{eg}$. \textbf{f -- j}, Respective normalized extinction obtained from the Fano model with exciton-plasmon coupling strength $\hbar\Omega_c=28$~meV.}
\label{figDetuning}
\end{figure*}

The parameters to obtain the extinction are known from the experiment or literature. For the exciton state \ket{e}, the population decay due to radiative recombination is $\hbar\Gamma_{ee} = \hbar\Gamma_{ee}^{r} \simeq 1$~meV \cite{Glazov2014,Moody2015}. The experimental FWHM linewidth is determined for the monolayer TMD region (Fig. 1c) as $\hbar\Gamma_{eg} \simeq 8$~meV, and we treat the broadening beyond the transform-limit of $1$~meV as pure dephasing. For the plasmon state \ket{p}, the experimental FWHM linewidth is $\hbar\Gamma_{pg} \simeq$ 180~meV. We neglect pure dephasing processes such that $\Gamma_{pg} = \Gamma_{pp}$ \cite{Sonnichsen2002}. The radiative contribution in the plasmonic state \ket{p} for the gold nanodisk is estimated to be $\simeq 46$ \%, such that $\hbar\Gamma_{pp}^{r} \simeq 82$~meV \cite{Sonnichsen2002}. The Rabi frequencies $\Omega_{eg,u}$ and $\Omega_{pg}$ are calculated using an illumination laser power $P = 1$~nW, illumination area $A \simeq$ \SI{1}{\micro\meter}$^{2}$, and the corresponding linewidth and resonance frequency. The Rabi frequency of excitons coupled to the plasmons $\Omega_{eg,c}$ is calculated using the enhancement factor, $F = 50$, and the coupled and uncoupled areas, $A_{c}$ and $A_{u}$, determined from the FDTD simulation shown in Fig.~\ref{figFDTD}. The coupling constant is estimated from the experiment to be $\hbar \Omega_{c} \simeq 28$~meV (see main text, Fig.~2e). A comparison between experimental differential reflectance spectra and extinction response from the model across different detunings is presented in Fig.~\ref{figDetuning} with very good agreement between the experiment and the model across a wide range of energy detunings. 

To determine the magneto-induced circular dichroism (CD), we calculate the extinction for different polarizations of light. Light with $\sigma^+$ ($\sigma^-$) polarization only couples to the $K$ ($K'$) valley of the monolayer exciton transition. These opposite valleys have different resonance energies $\omega_{eg}$ in the presence of an out-of-plane magnetic field according to the valley Zeeman splitting with the characteristic exciton $g$-factor of $4$ \cite{Srivastava2015,Wang2015b,Koperski2018,Wozniak2020,Foerste2020exciton,Deilmann2020,Xuan2020}. The optical response for $\sigma^+$ and $\sigma^-$ polarization is normalized by the sum of two optical responses in the calculation of CD, leading to the cancellation of the terms due to various reflecting surfaces in the device. Therefore, the calculated model CD values shown in main text are quantitative.

\clearpage

\section*{Supporting Note 3: Magneto induced chirality of the coupled system in differential reflection}

\begin{figure*}[htb!]
\centering
\includegraphics[scale=0.98]{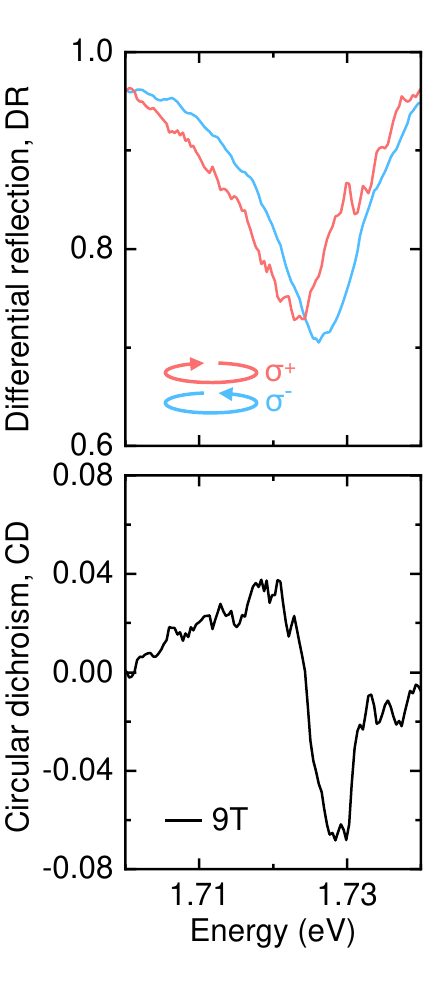}
\caption{\textbf{a}, Differential reflectance spectra of the coupled system around the Fano interference induced dip for different circular polarizations at a magnetic field of 9T. The transfer of the magneto induced chirality from the monolayer exciton to the coupled exciton-plasmon system is observed as an energy shift of the dip position for different polarization. The data of the two spectra for $\sigma^{+}$ and $\sigma^{-}$ detection lead to the CD spectrum shown in Fig.4c in the main text and in \textbf{b} here, for better comparability.}
\label{figDRchirality}
\end{figure*} 
